\documentclass[journal,10pt]{IEEEtran}
\ifCLASSINFOpdf
\else
   \usepackage[dvips]{graphicx}
\fi
\usepackage{url}

\usepackage{graphicx}
\usepackage{amsmath}
\usepackage{amssymb}
\usepackage{amsthm}
\usepackage{bbm}
\usepackage{subfigure}
\usepackage{bm}
\usepackage{bbold}
\usepackage[dvipsnames]{xcolor}
\usepackage{booktabs}
\usepackage{cite}
\allowdisplaybreaks[4]
\usepackage{multirow}
\usepackage{algorithm}
\usepackage{algorithmic}
\usepackage{stmaryrd}

\usepackage{xr}
\makeatletter
\newcommand*{\addFileDependency}[1]{
  \typeout{(#1)}
  \@addtofilelist{#1}
  \IfFileExists{#1}{}{\typeout{No file #1.}}
}
\makeatother

\begin{document}

\title{Implementation of an IEEE 802.11ax-Based Maritime Mesh Network in the Red Sea}

\author{Yingquan Li,~\IEEEmembership{Graduate Student Member,~IEEE}, Jiajie Xu,~\IEEEmembership{Member,~IEEE}, Narek Khachatrian, Mohamed-Slim Alouini,~\IEEEmembership{Fellow,~IEEE}
\thanks{Y. Li, J. Xu, N. Khachatrian, and M. S. Alouini are with the Computer, Electrical, and Mathematical Sciences and Engineering Division (CEMSE), King Abdullah University of Science and Technology (KAUST), Thuwal, 23955-6900, Kingdom of Saudi Arabia (email: yingquan.li@kaust.edu.sa, jiajie.xu.1@kaust.edu.sa, narek.khachatrian@kaust.edu.sa,  slim.alouini@kaust.edu.sa).}}

\maketitle

\begin{abstract}
In this article, we explore the limitations of satellite phones in meeting the communication needs of fishermen operating in the Red Sea. We propose AX-MMN, a maritime mesh network based on the IEEE 802.11ax standard, to address these shortcomings of satellite phones and outline AX-MMN's system architecture. To validate the performance of AX-MMN, we conduct extensive real-world experiments, demonstrating its potential to enhance maritime connectivity significantly. We also discuss the broader benefits of AX-MMN, particularly for fishermen in underdeveloped East African countries bordering the Red Sea, emphasizing its capacity to improve their communication capabilities and overall quality of life.
\end{abstract}

\begin{IEEEkeywords}
Maritime communications, IEEE 802.11ax, Red Sea, rural connectivity.  
\end{IEEEkeywords}

\IEEEpeerreviewmaketitle

\section{Introduction}\label{Sec:intro} 
As the global population exceeds 8 billion~\cite{Population}, the demand for natural resources, including marine resources, has reached unprecedented levels, leading to a substantial increase in human activity across the world’s oceans. In regions such as the Middle East and East Africa, over 100 million people engage in economic activities in the Red Sea, with many involved in fishing. Fishermen often rely on the islands in the Red Sea as temporary shelters during their fishing expeditions and travel between land and sea. However, despite there being more than 1,000 islands in the Red Sea~\cite{RedSea}, the majority are uninhabited and lack the communication infrastructure and terrestrial network coverage. As a result, these islands fail to provide adequate communication services, thereby hindering the fishermen’s ability to manage their operations and engage in leisure activities, while also compromising their safety and security during emergencies.

Satellite phones are the most commonly used solution for addressing the lack of maritime communication networks. However, satellite phones present several significant disadvantages:
\begin{itemize}
    \item High Cost: Satellite phones are expensive. Accessing satellite communication requires a device capable of transmitting and receiving signals within the specific frequency bands used for satellite communication, with sufficient transmission power and sensitivity. Most of the mobile phones available on the market only support cellular network communication within the coverage of terrestrial base stations and cannot connect directly to satellites. Therefore, one must purchase a dedicated device supporting corresponding frequency bands and protocols, typically costing thousands of dollars. Additionally, the charges for satellite phones are high, ranging from \$1 to \$2 per minute~\cite{Charge}, with monthly packages still costing hundreds of dollars. These costs are unaffordable for most fishermen.
    \item Limited Bandwidth: The data rates achievable with commercial satellite phones are between 2200 and 9600 bps~\cite{SatellitePhone}, equivalent to the speeds offered by GSM protocols during the 2G era. In contrast, terrestrial networks have already achieved 5G and are advancing toward 6G, and even in-flight WiFi services have become common. This highlights a significant lag in maritime communication networks, which are currently four generations behind terrestrial and aerial networks. Given the higher frequency of human activities on the ocean compared to the air, there is an urgent demand for new solutions that provide higher data rates, thereby replacing low-speed satellite phones.
    \item Lack of Networking Capabilities: Satellite phones are restricted to point-to-point communication and lack the capability for multi-device networking. This limitation hinders the use of applications like online meetings, social media, and multiplayer gaming, making satellite phones inadequate for delivering comprehensive communication services to fishermen at sea. Consequently, they fail to enable real-time collaboration among fishermen, such as sharing the location of fish schools or adjusting fishing plans. As a result, satellite phones do little to enhance the efficiency of maritime activities, increase fishermen’s income, or promote regional economic growth.
\end{itemize}
Given these shortcomings of the current maritime communication network, we propose a cost-effective maritime mesh network (AX-MMN) using the IEEE 802.11ax (Wi-Fi 6) standard to provide fishermen with affordable, high-speed, many-to-many communication services. A preliminary point-to-point test of this network was successfully conducted in the open waters near KAUST. The structure of this paper is as follows: Section~\ref{sec:system} introduces the system architecture of AX-MMN. Section~\ref{sec:exp} presents the conducted real-world experiments. Section~\ref{sec:benefit} provides the benefits that AX-MMN brings to marine communication. We conclude this article in Section~\ref{Sec:conclusion}.

\begin{figure*}
    \centering
    \includegraphics[width=\linewidth]{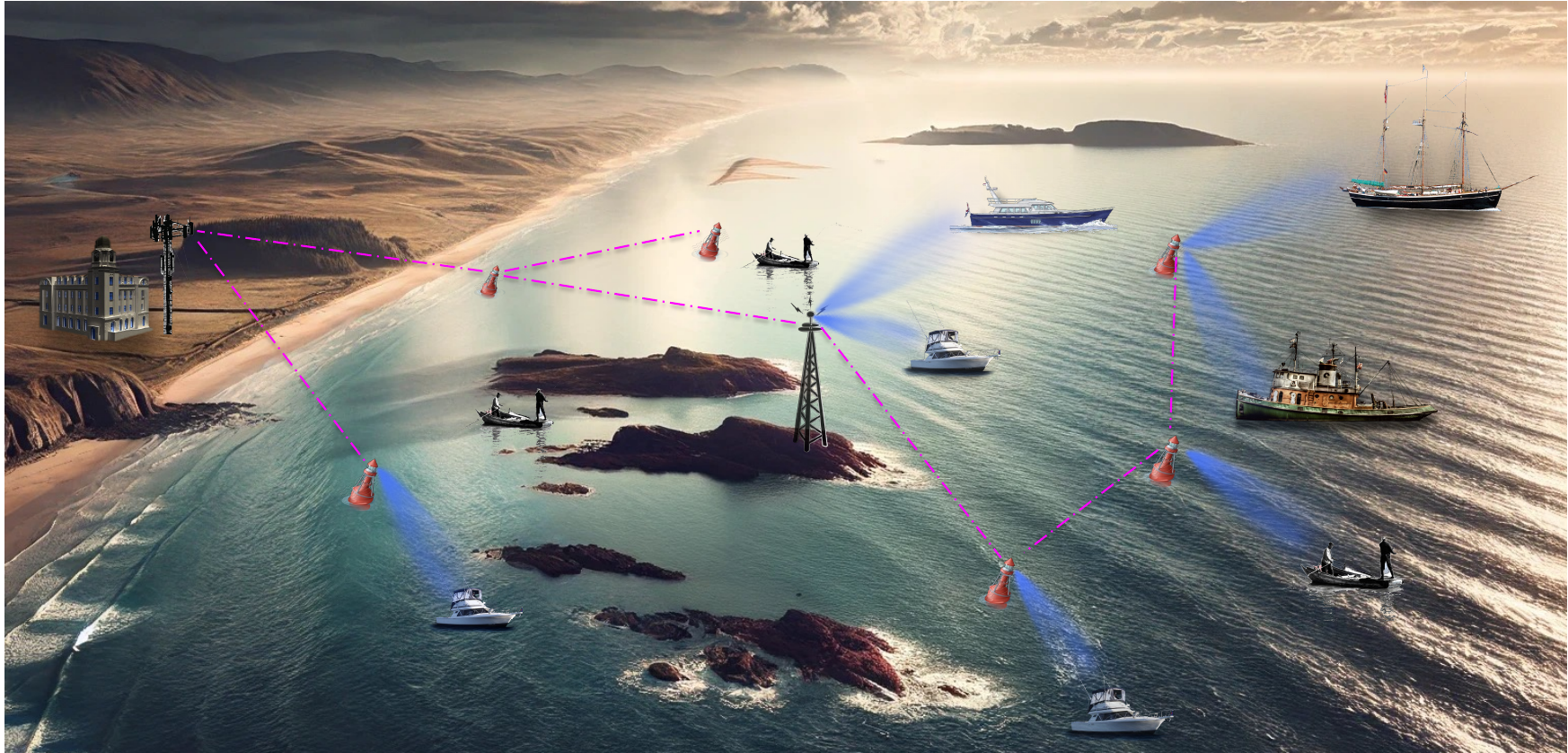}
    \caption{Depiction of the system architecture for AX-MMN.}
    \label{Fig:Architecture}
\end{figure*}
\section{System Architecture for AX-MMN}\label{sec:system}
In this section, we present the system architecture of our proposed maritime mesh network, designed to deliver cost-effective, high-speed, and comprehensive communication services to fishermen operating in the Red Sea. The mesh network leverages the protocol IEEE 802.11ax (Wi-Fi 6), using oceanic buoys and uninhabited islands as relay nodes to establish multi-hop links between fishermen and terrestrial networks. This design can effectively extend communication coverage to remote regions in the Red Sea where terrestrial or aerial network coverage is inadequate. 
\subsection{Network Topology}\label{subsec:NT}
We display the network topology of AX-MMN in Fig.~\ref{Fig:Architecture}. The network comprises various nodes, including fishermen’s mobile devices, relay stations on uninhabited islands (RS-UIs), buoys, and terrestrial base stations on the land. We use the RS-UIs and buoys as significant infrastructure to extend signal propagation distances and minimize the probability of communication outages. 

Compared to buoys, the number of RS-UIs is fewer. We recognize an island as a suitable place for deploying an RS-UI if it meets one of the following criteria: (1) fishermen commonly use this island as a shelter, and (2) this island's proximity to the shore facilitates the transportation of equipment. RS-UIs are less susceptible to displacement by ocean currents, ensuring robust operation. Owing to the sufficient space on the unmanned island, the RS-UI features a more complex structure and higher transmission power, enabling it to provide broader Wi-Fi coverage for fishermen in the surrounding area. 

Oceanic buoys, equipped with Wi-Fi 6 transceivers, are powered by a combination of solar energy and rechargeable backup batteries. To ensure positional stability, we use iron chains to anchor buoys, thereby restricting their movement to a limited range determined by the chain’s length. It is noteworthy that ``positional stability" does not imply that the buoys remain entirely stationary. Instead, it refers to the buoy's restricted mobility, as the anchoring mechanism prevents it from moving freely with ocean currents while allowing minor oscillations and limited displacements. The anchoring mechanism plays a significant role in the setup of the AX-MMN by preventing buoys from being overturned or displaced by external forces such as waves or currents. This ensures that the buoy consistently maintains its connection with neighboring nodes within the AX-MMN. The collaboration between buoys significantly enhances the robustness of the AX-MMN, enabling continuous and reliable communication in dynamic and challenging oceanic environments. Their low cost enables widespread deployment, addressing the stochastic access needs of nearby fishermen. The multi-hop network formed by multiple buoys extends the maritime mesh network's coverage to remote areas. RS-UIs and oceanic buoys function as relay nodes, facilitating connection between terrestrial networks and fishermen. 

Buoys exhibit greater positional variability than RS-UI, and restricting their movement entirely is challenging. However, limited buoys' movement due to mild ocean currents does not necessarily reduce system coverage. When slightly moving within an acceptable range, buoys can maintain stable mesh links with neighboring nodes, ensuring reliable relay performance. As long as it stays connected to nearby nodes, its movement may extend system coverage in the direction of the movement. Network coverage decreases, and the probability of outages increases under conditions of significant movement caused by strong currents, which weaken or disrupt connections between nodes. Furthermore, the IEEE 802.11ax standard can stand low-speed movement of connected nodes, which enhances the system's robustness to buoys' movement. Consequently, rather than investing substantial resources to completely restrict buoys' movement, the anchoring mechanism provides a cost-effective and practical solution. By controlling buoys' movement to a tolerable range, this approach ensures a minor impact on network coverage while maintaining ease of deployment and adaptability to maritime environments.
\subsection{Node Configuration}\label{subsec:NC}
The nodes in AX-MMN are categorized into terrestrial base stations, relay nodes, and mobile terminals. Terrestrial base stations are gateways situated on the coastline with access to existing terrestrial networks. These stations are equipped with high-power transceivers and serve as central processors for data aggregation, routing, and Internet access. However, their large size and high cost limit their deployment to the coastline rather than uninhabited islands. Consequently, only nodes located in offshore areas can directly communicate with these terrestrial base stations. 

Relay nodes extend network coverage by forwarding uplink and downlink packets, bridging communication between fishermen in remote oceans and the terrestrial network. These nodes are designed with minimal signal processing capabilities, but their low energy consumption and cost allow for widespread deployment on buoys and certain uninhabited islands, enhancing connectivity over vast ocean areas.

Mobile terminals are portable devices used by fishermen to access Wi-Fi networks. AX-MMN adopts Wi-Fi 6, which is backward compatible with Wi-Fi 5 and Wi-Fi 4~\cite{WiFi6compatible}. Therefore, fishermen can access the Internet using their smartphones and smartwatches, eliminating the need to purchase specialized equipment and significantly reducing communication costs. 
\subsection{Communication Protocol}\label{subsec:CP}
AX-MMN uses the IEEE 802.11ax standard, commonly known as Wi-Fi 6, to enhance data throughput and overall network performance. Operating in the 2.4 GHz and 5 GHz frequency bands~\cite{WiFi65G}, IEEE 802.11ax introduces several key features that make it particularly advantageous for maritime applications.

Firstly, IEEE 802.11ax employs Orthogonal Frequency Division Multiple Access (OFDMA), allowing multiple devices to share the same channel simultaneously~\cite{OFDMA}. This efficient spectrum utilization reduces latency and increases network capacity, which is significant in maritime environments where bandwidth is limited, but multiple fishermen need to connect to the network at the same time. 

Secondly, the implementation of Target Wake Time (TWT) in IEEE 802.11ax enables devices to schedule their wake times for transmitting or receiving data, thereby effectively saving battery life~\cite{TWT}. This feature is especially beneficial for buoys that rely on unstable solar power and operate without charging facilities in the open ocean.

Moreover, IEEE 802.11ax acts as a cost-effective alternative to satellite phones when delivering communication services. Fishermen can use their existing smartphones, laptops, and other common electronic devices to connect directly to the network, significantly reducing the entry cost for accessing communication services. Therefore, we adopt IEEE 802.11ax for constructing AX-MMN. Its advanced features not only enhance network performance and coverage but also provide fishermen with reliable and cost-effective communication services. This accessibility supports fishermen's daily activities and improves their safety, economic income, and overall quality of life.

AX-MMN aims to provide affordable and easily accessible communication services to fishermen through RS-UIs or buoys, rather than focusing on long-distance telecommunication using a single node. To achieve this goal, we adopt the Wi-Fi 6 standard for its compatibility with existing mobile terminals, avoiding the need for expensive specialized equipment. Although the propagation distance of Wi-Fi 6 signals is limited, the low cost of individual nodes allows for network coverage to be expanded by increasing the number of deployed nodes. A discussion of node costs is provided in Table~\ref{Tb:cost}. By increasing node density, AX-MMN can not only cover larger maritime areas but also maintain reliable performance.

In summary, the proposed system architecture for AX-MMN delivers long-range, reliable, low-cost, and comprehensive communication services within the Red Sea's maritime environment. By meeting the needs of maritime communication, the architecture exhibits a potential solution for enhancing communication capabilities across uninhabited islands and expansive ocean areas in the region.

\begin{figure}
    \centering
    \includegraphics[width=\linewidth]{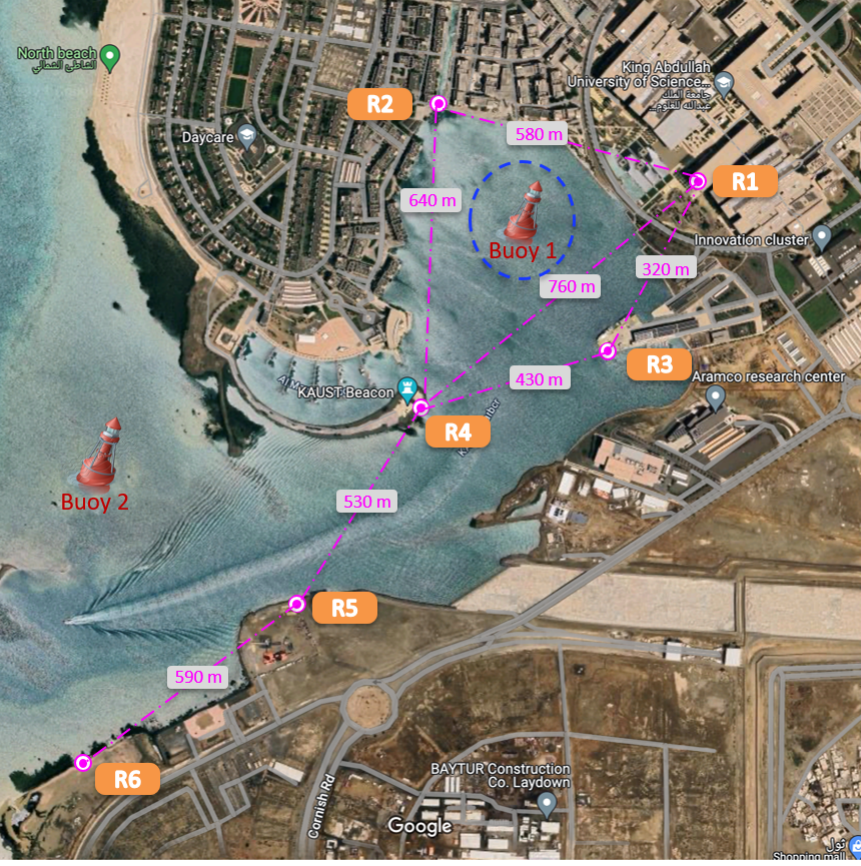}
    \caption{Photographs of the experimental site of AX-MMN.}
    \label{Fig:site}
\end{figure}
\section{Real-world Experiments}\label{sec:exp}
To validate the effectiveness of AX-MMN, we conduct large-scale marine experiments in the Red Sea, covering an area of $1$~${\text{km}}^2$. We deploy 8 nodes across the sea surface and coastline. We present the experimental site in Fig.~\ref{Fig:site}. Each node is equipped with 4 amplifiers and four 8~dBi antennas, specifically designed for outdoor environments. These large, directional antennas can significantly enhance the propagation distance. To further improve communication performance, we employ a 1024-QAM modulation scheme with a 160 MHz channel bandwidth. While the transmitted signal strength remains less than 500 mW in this system, the combination of high-performance amplifiers and optimized antennas effectively extends Wi-Fi coverage.

We place R1 on the roof of a four-story building, 18~m above ground level, to simulate a terrestrial base station. R1 is powered by a 220~V external power supply within the building. We place R2-R6 along the shore to simulate RS-UIs on uninhabited islands in the Red Sea, each with a height of at least 3~m. We employ a strategy that combines solar panels with a rechargeable power station to ensure a sustainable power supply for R2-R6. We separate the nodes to maintain a minimum distance of 300~m between any two, simulating the scenarios where nodes are sparsely distributed in a maritime environment. We place two buoys at sea, with one located inside the convex hull of R1-R6 to enhance network stability and the other placed outside this boundary to extend the network coverage. Each buoy is anchored 1.5~m above the sea and powered by solar energy. The maximum power consumption of nodes is 18~W, while the solar panels used have a maximum power output of 100~W. Under the sunlight conditions at KAUST, this is sufficient to achieve sustainability. Additionally, the rechargeable battery has a capacity of 768~Wh, ensuring that the nodes can operate continuously at night or under periods of insufficient sunlight. We installed routers supporting the IEEE 802.11ax standard on R1-R6 and the two buoys to provide Wi-Fi services. The router operates on 2.4~GHz and 5~GHz frequency bands, achieving a maximum data rate of 2976~Mbps: 574~Mbps on the 2.4~GHz band and 2402~Mbps on the 5~GHz band. We equipped each router with four 5~dBi omnidirectional antennas. The router features a dual-core 1.3~GHz CPU and 2~GB of memory, enabling it to form a mesh network with up to 256 devices~\cite{Router}. The router measures $175\times125\times34$~${\text{mm}}^3$ and weighs 255~g, with a power consumption of 12~W. Its compact size facilitates easy installation and material cost savings during deployment. Furthermore, the low power consumption allows for sustainable operation with solar panels, making it suitable for large-scale deployment in remote and extreme environments, such as uninhabited islands and buoys in the Red Sea. 

\begin{figure}
    \centering
    \includegraphics[width=\linewidth]{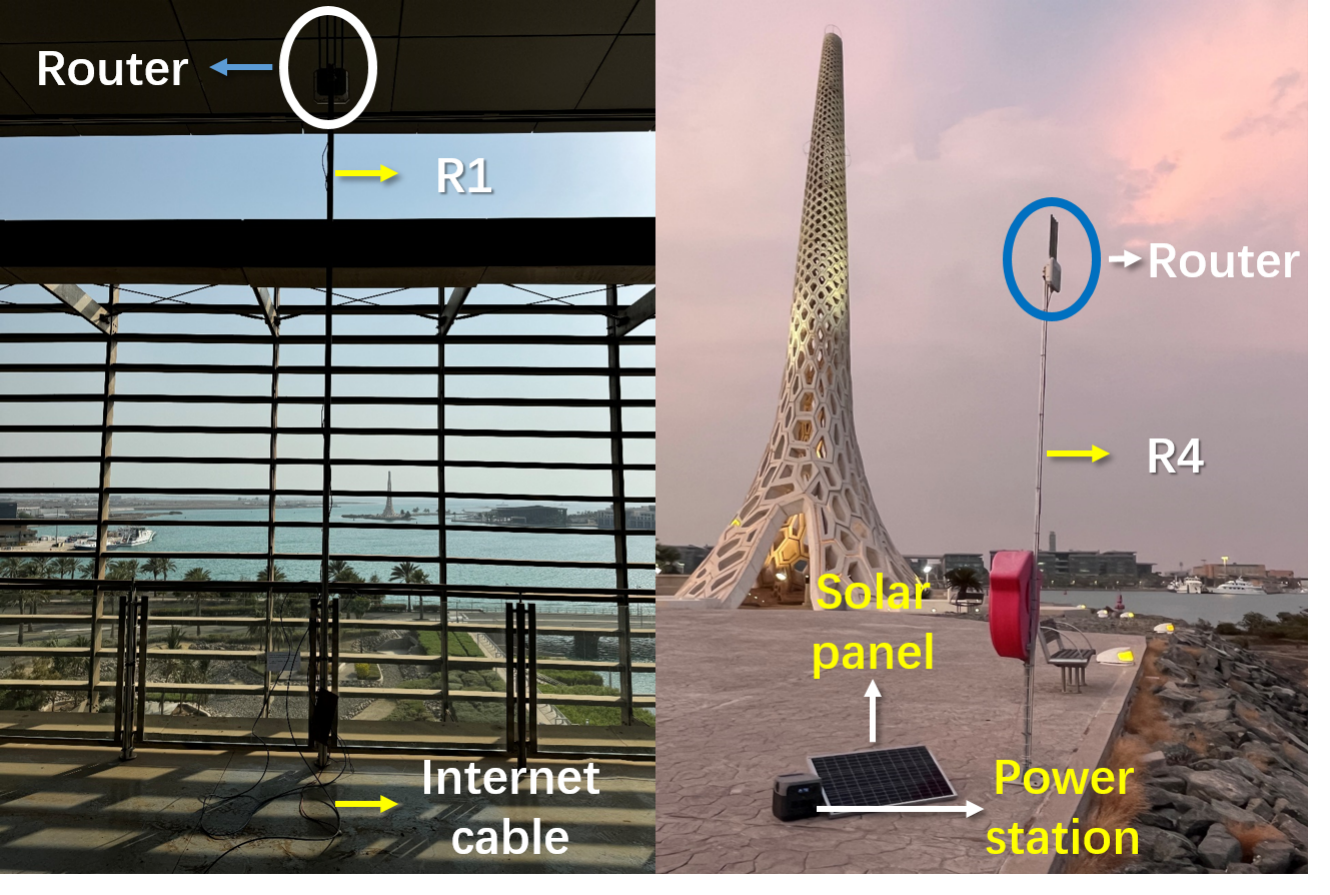}
    \caption{Photographs of the experimental setup for R1 and R4.}
    \label{Fig:Setup}
\end{figure}
\begin{figure}[!t] 
    \centering 
    \subfigure[Data Rates Measured Near R3.]{
    \centering
    \includegraphics[width=.7\linewidth]{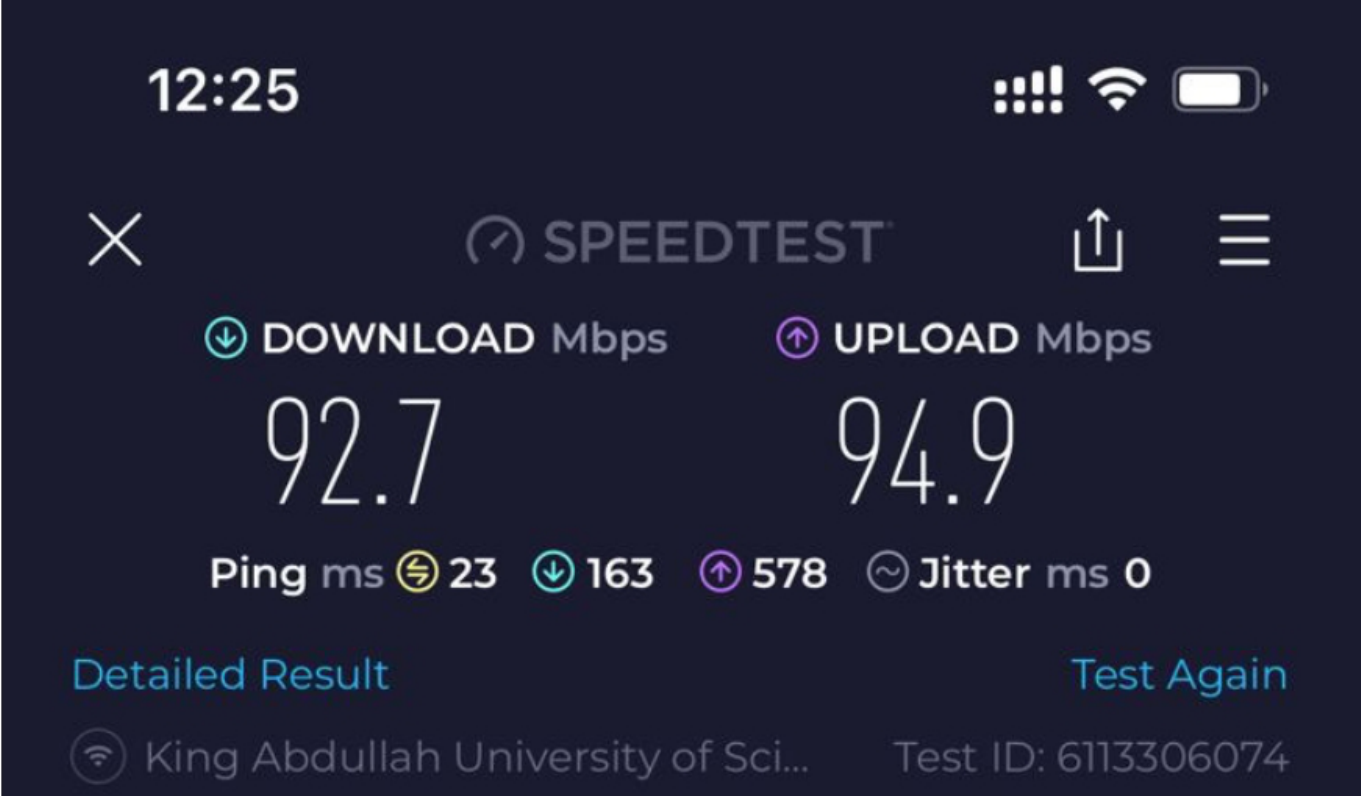} 
    \label{Subfig:R6}
    }
    \subfigure[Data Rates Measured Near R6.]{
    \centering
    \includegraphics[width=.7\linewidth]{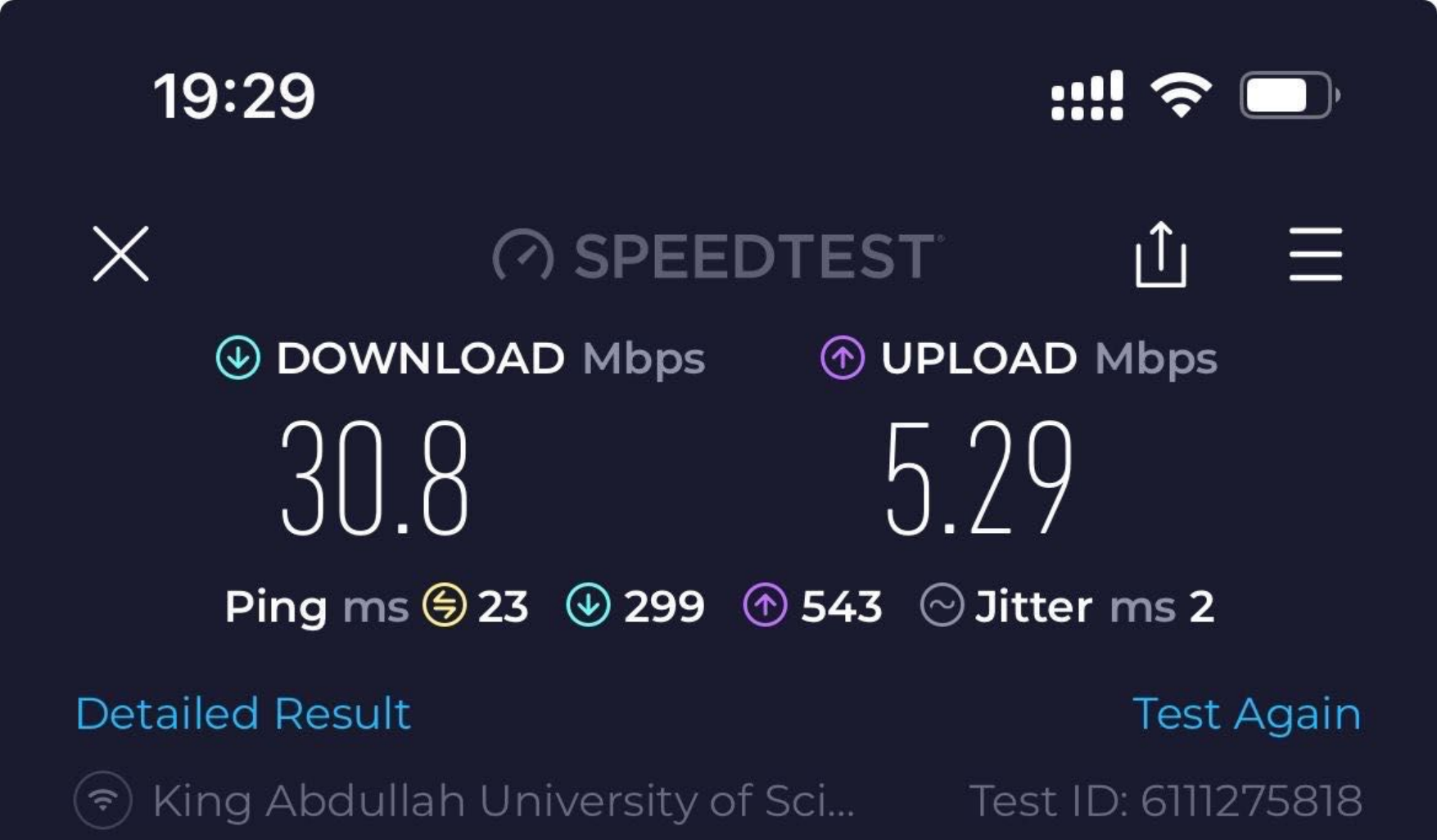} 
    \label{Subfig:R3}
    }
    \caption{Photograph of data rates measured in the maritime region surrounding R3 and R6.}
    \label{Fig:DataRate}
\end{figure}
We present the photography for R1 and R4 in Fig.~\ref{Fig:Setup}. During the experiment, we conducted sea cruises among nodes to simulate the common random and mobile access scenarios experienced by fishermen. We evaluated the performance of AX-MMN under realistic conditions by measuring uplink and downlink data rates using smartphones. Field tests show that internet access is unavailable on the sea surface near R5-R6 with only R1 deployed. However, with the addition of buoys and other nodes, the internet coverage can be extended throughout the experimental sea area. We use three types of terminals to evaluate the performance of AX-MMN: a smartwatch, a mobile phone, and a laptop. We tested data rates near R3 and R6 using a mobile phone, which were initially unconnected. The results of these tests are presented in Fig.~\ref{Fig:DataRate}. Since R3 can establish a direct connection to R1, we recorded a higher data rate of 90 Mbps. In contrast, at R6, the device accessed the Internet by connecting to R1 via other nodes or buoys serving as relays. Despite the 2 km distance between R6 and R1, we recorded a downlink data rate of 30 Mbps and an uplink data rate of 5 Mbps. These data rates facilitate various applications for fishermen. They can engage with social media, maintain community connections, and seek assistance during emergencies. We conducted additional tests near Buoy 2, measuring stable download data rates of approximately 20 Mbps at two coastal locations. The results demonstrate the system's capability to support communication services such as video calls, streaming, and real-time weather updates. For fishermen, this connectivity enhances fishermen's safety and communication with onshore facilities, confirming the system's reliability in coastal and offshore environments. At R4, one user utilizing a mobile phone conducted a video conference with another user employing a laptop through AX-MMN, thereby demonstrating the system's superior end-to-end performance. A recording of this demonstration is available in \cite{Video1}. The capability to stream music and videos enhances leisure activities for fishermen and helps mitigate the psychological stress associated with prolonged periods at sea. The high data rates can enable video calls, allowing fishermen to communicate uninterruptedly with family members and promptly handle private matters. This ensures they remain connected to their personal lives while at sea, even under challenging conditions. Reliable connections facilitate the use of Internet of Things (IoT) devices for the monitoring and management of onboard equipment, thereby enhancing the efficiency of fishing operations. In addition, enhanced connectivity supports real-time collaboration with nearby fishermen through online platforms, leading to cooperative decision-making and information exchange~\cite{ShipAccident}. 

\begin{table}
\centering
\caption{The expenses associated with conducting the real-world experiments.}
\begin{tabular}{|c|c|c|c|c|}
\hline
No. & Item          & Unit & Unit Price (USD) & Total Cost (USD) \\ \hline
1   & Router        & 8    & 78.79            & 630.32           \\ \hline
2   & Power Station & 5    & 110.06           & 550.30           \\ \hline
3   & Solar Panels  & 7    & 42.02            & 294.14           \\ \hline
4   & Misc.         & /    & /                & 14.00            \\ \hline
    &               &      &                  & 1488.76          \\ \hline
\end{tabular}
\label{Tb:cost}
\end{table}
To evaluate the networking distance between nodes in AX-MMN, we turned on R1, R4, and R6 while deactivating all other nodes. A data rate of 30~Mbps was measured near R6. A recording of the demonstration is available in \cite{Video2}. This experiment demonstrates that R6 can access the terrestrial network by using a relay located 1 km away, thereby showcasing the extensive coverage capabilities of AX-MMN. In this sparse network, Wi-Fi 6 signals can maintain satisfactory data rates even after a transmission distance of 2~km. The ability of nodes to establish long-distance connections significantly enhances AX-MMN's coverage expansion, enabling reliable communication services over a broad maritime area with a minimal number of nodes. Given the low cost of the nodes, AX-MMN is well-suited for maritime communication where infrastructure is limited. We present the costs associated with the experimental system in Table~\ref{Tb:cost}. Notably, our proposed maritime mesh network can provide high-speed, stable, and multi-application communication over a $1$~${\text{km}}^2$ sea area at a cost comparable to that of a satellite phone. Despite the highly variable wireless channels caused by frequent storms, turbulence, and high temperatures in the Red Sea near Jeddah~\cite{JeddahWeather}, the system has operated autonomously and reliably for two months, demonstrating its practicality and robustness. During this period, all nodes except R1 are solely powered by solar energy, showcasing the system's sustainability. This low-maintenance design makes it suitable for less developed regions, such as the East African nations along the Red Sea coast. In summary, these experimental results demonstrate that AX-MMN significantly improves fishermen’s quality of life, bridges the digital divide, and integrates them into the broader 6G era by expanding access to information, entertainment, and essential services.

\section{Benefits of AX-MMN for Marine Communication}\label{sec:benefit}
Reliable maritime communication is significant for enhancing the efficiency and safety of fishermen operating in remote sea environments. AX-MMN provides several key benefits that can enhance the communication capabilities of fishermen. By delivering low-latency communication, high bandwidth, and seamless integration with terrestrial networks, AX-MMN fulfills the specific requirements of maritime scenarios and addresses the challenges encountered by fishermen. In this section, we discuss each of these benefits in detail, demonstrating how they contribute to more effective, reliable, and comprehensive communication services for fishermen operating in remote sea environments.
\subsection{Low Latency}
The low latency provided by AX-MMN enhances the system performance of maritime communication, particularly for fishermen operating in the challenging environment of the Red Sea. Low-latency communication is essential for high-precision GPS tracking systems, enabling fishermen to accurately determine their positions and locate fish schools in real time. This capability can result in optimizing fishing operations and making decisions, such as adjusting travel direction or schedules. In cooperative fishing scenarios, where communication between nearby fishermen is necessary, low latency ensures that information is exchanged swiftly, allowing for coordinated actions and reducing the risk of delays that could hinder operational efficiency.

Furthermore, the Red Sea's narrow channels pose a frequent risk of ship collisions~\cite{ShipAccident}, typically due to the lack of low-latency communication methods between vessels. AX-MMN’s ability to deliver low-latency communication can significantly mitigate this risk, enhancing the safety of navigation and preventing potential accidents. Overall, the low latency provided by AX-MMN is a significant factor that enhances both the efficiency and safety of fishing operations, ultimately contributing to the well-being and economic promotion of the less developed community near the Red Sea.
\subsection{Cost-Effectiveness} \label{subsec:CEff}
Before the widespread adoption of maritime communication technology, cost is a crucial consideration, particularly for less developed East Africa around the Red Sea. In designing and deploying AX-MMN, we prioritize cost-effectiveness, ensuring that the network is both affordable and accessible to the majority of fishermen while maintaining high performance.

Compared to existing satellite and offshore networks, AX-MMN provides stands out by enabling cost-effective, many-to-many communication over expansive ocean areas. Satellite phones can support marine communication but come with significant drawbacks, including high costs, limited bandwidth, and a lack of support for Internet services like video streaming. Similarly, offshore networks are constrained by their coverage, rendering them unsuitable for users operating in remote maritime regions.

AX-MMN is designed to overcome these limitations by providing wide-area Internet access at a low cost. Its ability to support streaming services and mesh networks across maritime users makes it more versatile than traditional marine communication solutions. To achieve this, AX-MMN leverages the Wi-Fi 6 standard, chosen for its backward compatibility with existing mobile devices. This compatibility ensures that fishermen can use their current mobile devices without incurring additional expenses. Although Wi-Fi 6 signals have a limited propagation distance, the cost-effectiveness of AX-MMN enables the network's coverage to be extended rapidly by deploying additional nodes. AX-MMN presents a scalable and cost-effective solution for extending connectivity to the open ocean, playing a role in bridging the digital divide in unconnected regions.

By balancing network costs with enhanced communication capabilities, AX-MMN addresses the connectivity needs of maritime users, particularly in underdeveloped regions. Its scalability and affordability make it a suitable solution for maritime communication infrastructure, showcasing reliable and flexible services to improve the quality of life for fishermen.

Moreover, cost-effectiveness can extend network coverage to fishermen in less developed nations and unconnected remote sea environments. By deploying low-cost buoys across major areas where fishermen operate, AX-MMN avoids coverage gaps and ensures comprehensive connectivity, regardless of location. The network's reliance on solar-powered nodes also effectively reduces maintenance costs, reduces the financial requirement for communication operators, and prevents potential service interruptions due to high operating expenses. By keeping costs manageable, communication operators can ensure the network's long-term sustainability, providing reliable communication services to fishermen over time. Furthermore, the use of solar energy aligns with global sustainable development goals by minimizing the environmental impact on marine ecosystems.

In conclusion, the cost-effectiveness of AX-MMN leads to making communication services accessible, scalable, sustainable, and economically friendly for fishermen. Prioritizing affordability allows us to address the diverse needs of less developed communities, ensure the network's long-term success and sustainability, and deliver efficient, stable communication solutions for fishermen.
\subsection{Coexistence with existing maritime communication networks}\label{subsec:coe}
AX-MMN can coexist with existing maritime communication networks without interference. Traditional maritime communication systems, such as VHF radios, operate in the frequency range between 156 and 174~MHz, which is commonly used for communication between ships, shore stations, and sometimes aircraft. However, AX-MMN, which operates on the 2.4~GHz and 5~GHz bands as defined by the IEEE 802.11ax standard, avoids interference with these legacy communication systems. This ensures that AX-MMN can provide additional connectivity without disrupting critical communication services already in place.

By leveraging these higher-frequency bands, AX-MMN effectively enhances the communication capabilities in maritime environments, particularly by providing more robust and higher-capacity services such as internet access, data transfer, and streaming services. This compatibility allows AX-MMN to complement existing systems rather than replace them, making it an efficient solution for expanding communication coverage in remote areas of the ocean.

AX-MMN introduces additional frequency bands to the maritime communication network, assisting existing systems to enhance the overall robustness of marine communications. By operating on the 2.4~GHz and 5~GHz bands, AX-MMN provides an alternative option, which is beneficial when the channel quality in one frequency band deteriorates. In such scenarios, maritime users can maintain communication through other available frequency bands to ensure uninterrupted connectivity. This multi-band network can significantly contribute to the safety of maritime users by providing an additional layer of redundancy, allowing them to stay in contact with the outside world in challenging maritime environments. To sum up, the coexistence of AX-MMN with existing maritime communication technologies enhances network reliability and ensures uninterrupted connectivity, thereby improving safety and operational efficiency at sea.
\subsection{Prospects of integration with non-terrestrial networks (NTN)}\label{subsec:NTN}
NTN represents a promising approach to extending connectivity by integrating satellite, high-altitude platforms (HAPS), drones, and other non-terrestrial platforms into traditional communication systems. NTN is particularly significant for mitigating the digital gap across remote and unconnected regions, such as maritime environments where terrestrial infrastructure is insufficient or unavailable. By leveraging satellites and aerial platforms, NTN can provide wide-area coverage and enable broadband communication over long distances, making them essential for addressing the challenges of maritime connectivity.

AX-MMN is designed as a cost-effective and easily accessible maritime network, featuring a flexible architecture and reliable performance that make it feasible to integrate with NTN. AX-MMN uses the IEEE 802.11ax standard, which supports interoperability with a wide range of existing devices and communication frameworks. This adaptability allows AX-MMN to function as an important local extension of NTN, utilizing NTN’s backbone to relay long-distance traffic load while providing affordable and reliable last-mile connectivity through its mesh network. The nodes in AX-MMN can act as relays, facilitating the extension of NTN-provided connectivity to end-users, such as fishermen, in a cost-effective and efficient manner. This potential for integration highlights AX-MMN’s versatility and underscores its value as a fundamental component of global communication networks in the future.

The integration of NTN with AX-MMN can enhance connectivity stability for large vessels, addressing communication disruptions caused by the metal structure of the ship and the varying orientation of antennas. NTN’s ability to provide near-global coverage and support seamless connectivity between remote areas and ground networks is beneficial for maritime applications. Moreover, NTN’s HAPs can assist in ensuring continuous communication by offering elevated points that reduce the impact of local disruptions and fading signals. This integration not only improves link availability but also increases network resilience, addressing the challenge of sparse node distribution in AX-MMN. Consequently, the integration of NTN with AX-MMN holds potential for enhancing the robustness and scalability of maritime networks, providing affordable, continuous, and high-quality communication services over vast ocean areas.

The integration of AX-MMN with NTN combines the strengths of both systems. NTN can deliver extensive coverage and high-speed connectivity, while AX-MMN can expand this coverage to localized users, ensuring reliable and scalable communication. This integration enhances the reliability of global networks and extends connectivity to maritime scenarios, thereby improving the quality of life for underserved populations, such as fishermen in the Red Sea. This mutual practicality highlights the benefits of integrating the two systems.

\section{Conclusion}\label{Sec:conclusion}
In this article, we have examined the limitations of current satellite phones in providing communication services to fishermen operating in the Red Sea. To overcome the challenge of effective maritime communication, we have proposed a maritime mesh network based on the IEEE 802.11ax standard and outlined its system architecture. The effectiveness of AX-MMN has been validated through extensive real-world experiments, yielding results that demonstrate its capability to enhance maritime connectivity. Furthermore, we have discussed this network's significant benefits to fishermen, particularly in the less developed East African regions surrounding the Red Sea.

\bibliography{IEEEabrv,ref}
\bibliographystyle{IEEEtran}

\section*{biographies}
\begin{IEEEbiographynophoto}
{Yingquan Li} is a Ph.D. candidate at King Abdullah University of Science and Technology (KAUST), Thuwal, Saudi Arabia. 
\end{IEEEbiographynophoto}

\begin{IEEEbiographynophoto}
{Jiajie Xu} is a Postdoctoral Fellow at KAUST, Thuwal, Saudi Arabia.
\end{IEEEbiographynophoto}

\begin{IEEEbiographynophoto}
{Narek Khachatrian} is a Ph.D. student at KAUST, Thuwal, Saudi Arabia. 
\end{IEEEbiographynophoto}

\begin{IEEEbiographynophoto}
{Mohamed-Slim Alouini} (Fellow, IEEE) is a distinguished professor at KAUST.
\end{IEEEbiographynophoto}

\end{document}